\documentclass[]{proarticle}
\usepackage{multicol}
\usepackage{graphicx}
\usepackage{amssymb}
\usepackage{epsfig}

\usepackage{latexsym}
\usepackage{amsmath}
\usepackage{multibox}
\usepackage{verbatim}
\usepackage{mathrsfs}

\usepackage{amssymb,amsmath}
\usepackage{amsthm}
\usepackage{amscd}
\usepackage{amssymb}
\usepackage{amsfonts}
\usepackage{url}
\usepackage{slashed}
\usepackage[latin1]{inputenc}
\usepackage[english]{babel}
\usepackage{psfrag}

\newcommand{\beq}{\begin{equation}}
\newcommand{\eeq}[1]{\label{#1}\end{equation}}
\newcommand{\bea}{\begin{eqnarray}}
\newcommand{\eea}[1]{\label{#1}\end{eqnarray}}

\begin{document}
\title{Spacetime Dimensionality from de Sitter Entropy}
\author{Arshad Momen$^a$ and Rakibur Rahman$^b$}
\maketitle
\address{$a$)  Department of Theoretical Physics, University of Dhaka, Dhaka 1000, Bangladesh\\[5pt]
$b$) Max-Planck-Institut f\"ur Gravitationsphsyik (Albert-Einstein-Institut), Am M\"uhlenberg 1, 14476 Potsdam-Golm, Germany}
\eads{amomen@univdhaka.edu, rakibur.rahman@aei.mpg.de}
\begin{abstract}
We argue that de Sitter universes with a small cosmological constant are entropically favored to have three spatial dimensions.
The conclusion relies on the causal-patch description of de Sitter space, where fiducial observers experience local thermal
equilibrium up to a stretched horizon, on the holographic principle, and on some assumptions about the nature of gravity and the
constituents of Hawking/Unruh radiation.
\end{abstract}


\begin{multicols}{2}
\section{Introduction}\label{sec1}
%
The fact that the observable Universe has three large dimensions of space defies explanations other than anthropic~\cite{4D}.
This issue is sharpened by String Theory, which allows a humongous multitude of universes with various dimensions~\cite{Landscape}.
One may wonder whether the Anthropic Principle is the only way to understand the observed spacetime dimensionality.

The standard view of cosmology holds that the Universe began with an epoch of inflation, during which spacetime geometry was approximately
de Sitter~\cite{Inflation}. One may present particular frameworks in which inflation
is one among various competing cosmological scenarios, whose relative probabilities of creation can be quantified~\cite{causalpatch,Sorbo}.
In the braneworld context in String Theory, it can be argued that the quantum creation of inflationary universes prefers one similar
to our early Universe~\cite{Sarangi}. String gas cosmology may also shed light on how only three (or less) spatial dimensions could
have grown into a macroscopic size~\cite{BV}. Other attempts to understand the  dimensionality of spacetime can be made
in the context of brane gas dynamics~\cite{Others} or by invoking some entropic principle~\cite{Others1}.

On the other hand, the late phase of our Universe is asymptotically de Sitter with a small cosmological constant~\cite{dS}.
One may ask whether this ultimate phase can be (partly) understood by an entropic principle:
the final spacetime configuration must have maximum entropy for a given amount of energy.
In this note we will argue, under certain fairly justifiable assumptions (to be spelled out as we proceed), that an asymptotically
de Sitter universe with a small cosmological constant is entropically favored to have three spatial dimensions.
Our regime of interest for the possible number of spatial dimensions, $d$,
is $2\leq d\leq10$. Such a restriction follows if one assumes that gravity is described by General Relativity in the infrared,
and that the underlying theory of quantum gravity yields Supergravity as some low-energy approximation.
%
\section{de Sitter Space \& Entropy Thereof}\label{sec2}
%
We will use the natural units: $c=\hbar=k_\text B=G=1$. In $(d+1)$ spacetime dimensions, this sets to unity the Planck length,
$l_\text P\equiv\sqrt[d-1]{G\hbar c^{-3}}$, and the Planck mass, $M_\text P\equiv\hbar c^{-1}l_\text P^{-1}$, which however may
be carried around for the sake of clarity.

Let us write down the ($d$+1)-dimensional de Sitter metric in the \emph{static} coordinates: \beq ds^2=-(1-H^2r^2)\,
dt^2\,+\,\frac{dr^2}{1-H^2r^2}\,+\,r^2\,d\Omega^2_{d-1},\eeq{metric} where $d\Omega^2_{d-1}$ is the line element
on $S^{d-1}$, and the Hubble parameter $H$ is related to the (positive) cosmological constant as:
$\Lambda=\tfrac{1}{2}d(d-1)H^2$. The apparent singularity at $r=1/H$ is a coordinate artifact. One can
analytically extend the metric to a geodesically complete spacetime of constant curvature with topology $S^d\times R^1$,
where $r=0$ represents antipodal origins of polar coordinates on a $d$-sphere. However, no single observer can access the
entire de Sitter space: an observer at $r=0$ experiences the presence of an event horizon at a distance $r=1/H$. The
``causal patch'' of the observer is the region which is in full causal contact with her, namely $0\leq r\leq 1/H$. The horizon
is observer-dependent in that \emph{any} observer following a time-like geodesic can be chosen to be at $r=0$, and two such
observers will belong to different causal patches. While the isometry group for de Sitter space is $SO(d+1,1)$, the
manifest symmetries of the causal patch are $SO(d)$ rotations plus translation in $t$. The remaining $d$ compact and $d$
non-compact generators displace an observer from one causal patch to another.

In what follows we will restrict all attention to a \emph{single} causal patch,~\`a la~\cite{causalpatch,dSsymmetry}.
As regions that are out of causal contact with a particular observer have no operational meaning to her, the observer should
consider the physics inside her horizon as complete, without making reference to any other region. Without loss of generality,
this we can choose to be the ``southern''  causal patch, where the Killing vector $\partial_t$ is time-like \emph{and}
future-directed, so that time evolution is well defined. We imagine that the causal patch is filled with ``fiducial observers''
(FIDOs), each of whom is at rest relative to the static coordinate system, i.e. each is located at a fixed $r$ and fixed values
of the angular variables. The only geodesic observer is the FIDO at $r=0$, whom we call the ``principal investigator'' (PI).
The PI can send a request to any other FIDO to perform certain \emph{local} measurements and report the results, which the PI
will eventually receive after waiting for a finite amount of time.

The PI at $r=0$ detects a thermal radiation with a temperature $T_{\text{GH}}=H/2\pi$$-$the
Gibbons-Hawking temperature of de Sitter space~\cite{GH}. More generally, a FIDO at a radial position $r$,
whose Killing orbit has a proper acceleration $\mathfrak{a}=H^2r/\sqrt{1-H^2r^2}$, detects a thermal bath with an effective
\emph{local} temperature~\cite{temperature}: \beq T(r)=\frac{1}{2\pi}\sqrt{H^2+\mathfrak{a}^2}=\frac{H}{2\pi\sqrt{1-H^2r^2}},
\eeq{toltemp} which is just the Gibbons-Hawking temperature multiplied by a Tolman factor~\cite{Tolman}. Using an Unruh-like
detector~\cite{Unruh}, the FIDO can indeed discover a thermal radiation with the temperature $T(r)$. This effect is real,
and can also be understood as pure Unruh effect associated with Rindler motion in the global embedding Minkowski space~\cite{embedding}.
The local temperature, however, blows up at the horizon $r=1/H$. A way to regularize this divergence is
to consider a ``stretched horizon''~\cite{stretchedhorizon}, that extends from the mathematical horizon (by some
Planck length) up to some $r_\text{c}<1/H$. The thickness may well be a physical
reality, originating possibly from quantum fluctuations~\cite{Maggiore}. The temperature measured at
the stretched horizon is then large but finite:
\beq T_\text{c}\equiv T(r=r_\text c)=\frac{H}{2\pi\sqrt{1-H^2r_\text{c}^2}}<\infty,\eeq{cutoff}
which sets a cutoff value for the temperature. It is natural to identify the cutoff with the Planck scale. In our case
it will indeed turn out that $T_\text c\sim M_\text P$.

A global notion of temperature is not meaningful in curved spacetime~\cite{Tolman}. Instead, one may need to introduce
operationally meaningful local concepts of temperature and thermal equilibrium~\cite{LTE}. It was shown in Ref.~\cite{temperature}
that the unique invariant \emph{locally Minkowskian} state of quantum fields in de Sitter space has exactly the
temperature given by~(\ref{toltemp}). We will therefore consider  only \emph{local thermal equilibrium} with temperature
$T(r)$ of the physical degrees of freedom (DOF) accessible to a FIDO at a radial position $r$.
What DOFs does the thermal radiation contain, i.e., what are the constituents of Hawking/Unruh radiation?
Postponing justification until later, let us assume that the accessible number of DOFs, $\mathfrak D$, does not depend
on $r$.

The PI can ask a FIDO at radial position $r$ to measure the local entropy density, and receive the result
\beq
\sigma(r)=\left(\frac{d+1}{d}\right)\mathfrak D\,a(d)\,[T(r)]^d,
\eeq{density}
where $a(d)$$-$the radiation constant per DOF in $d$ spatial dimensions$-$is given by
\beq
a(d)=\frac{\omega(d)}{(2\pi)^{d}}\,\zeta(d+1)\,
\Gamma(d+1),\quad \omega(d)\equiv\frac{2\pi^{d/2}}{\Gamma\left(\tfrac{d}{2}\right)},
\eeq{radcons}
$\omega(d)$ being the surface area of the boundary of a unit $d$-ball. The PI can take the volume integral
of~(\ref{density}) to compute the total thermal entropy of the causal patch: \beq S=\omega(d)
\int_0^{r_\text c}\frac{dr\,r^{d-1}}{\sqrt{1-H^2r^2}}\;\sigma(r).\eeq{a4} This integral can be expressed in terms
of hypergeometric functions by the variable redefinition: $x\equiv H^2r^2$, and the use of the
integral representation~\cite{abramowitz}: \beq \int_0^z\frac{dx\,x^{\alpha-1}}{(1-x)^{1-\beta}}=\frac{z^\alpha}{\alpha}(1-z)^\beta\,
_2F_1(\alpha+\beta,1;\alpha+1;z),\eeq{hypergeo} which holds for $\mathfrak{R}(\alpha)>0$. Thereby the PI finds that the total entropy
amounts to
\bea S=\left(\frac{d+1}{d^2}\right)\mathfrak D\,\omega(d)\,a(d)\left(\frac{\sqrt{1-\epsilon}}{2\pi}\,\right)^d
\left(\frac{1}{\sqrt \epsilon}\right)^{d-1}\nonumber\\ \times\,{}_2F_1\left(\tfrac{1}{2},\,1;\,\tfrac{d+2}{2};\,1-\epsilon\right),
\eea{a6}
where $\epsilon$ is a positive number defined as \beq \epsilon\equiv 1-H^2r_\text{c}^2=\left(\frac{H}{2\pi T_\text{c}}
\right)^2.\eeq{epsdef} Note that $\epsilon\ll 1$, because we consider $H/2\pi$ to be much smaller than $T_\text c\sim M_\text P$,
which is necessary for a semiclassical treatment to be valid. The total entropy~(\ref{a6}), which clearly diverges in the
limit $\epsilon\rightarrow 0$, is rendered large but finite by the stretched horizon.

Now, there is an entropy associated with de Sitter horizon, known as the Gibbons-Hawking entropy~\cite{GH}, which is
$\tfrac{1}{4}$ of the horizon area, $A$, in Planck units: \beq S_{\text{GH}}=\frac{1}{4}\,\frac{A}{l_\text{P}^{~d-1}}
=\frac{1}{4}\,\omega(d)\left(\frac{M_{\text{P}}}{H}\right)^{d-1}.\eeq{ghe}
To see its possible connection with the total entropy~(\ref{a6}) of the causal patch, we note that the hypergeometric
function appearing in the latter can be written as~\cite{abramowitz}: \beq _2F_1\left(\tfrac{1}{2},\,1;\,\tfrac{d+2}
{2};\,1-\epsilon\right)\equiv\left(\frac{d}{d-1}\right)\,\left[\,1+\delta(d,\epsilon)\,\right],\eeq{a9} where the function
$\delta$ is such that it vanishes in the limit $\epsilon\rightarrow0$. Thanks to Eq.~(\ref{a9}) and some properties of the
gamma function, one can rewrite the entropy~(\ref{a6}) as \beq S=\omega(d)\left[\frac{\mathfrak{D}\,\Gamma\left(
\tfrac{d+3}{2}\right)\zeta(d+1)}{(d-1)(\sqrt{\pi}\,)^{d+3}}\right]\left(\frac{T_\text{c}}{H}\right)^{d-1}+~...~,\eeq{a7}
where the ellipses stand for subleading terms. Their $H$-dependencies differ from that of the leading term, which mimics
the area law~(\ref{ghe}) for de Sitter entropy.

Now we invoke the holographic principle, which entails that the leading term in~(\ref{a7}) should be identified with the
Gibbons-Hawking entropy~(\ref{ghe}) of the de Sitter horizon~\cite{HMS}\footnote{Defining the entropy of the causal patch
as~(\ref{a4}) and identifying it with the Gibbons-Hawking entropy~(\ref{ghe}) has also been suggested by N.~Kaloper in a
talk titled ``Inflation and leaky cans''. We thank L.~Sorbo for pointing this out.}.
This relates the cutoff temperature
$T_\text c$ and the Planck mass $M_\text P$ in the following way: \beq \frac{T_\text c}{M_\text P}
=\left[\frac{(d-1)(\sqrt{\pi}\,)^{d+3}}{4\,\mathfrak{D}\,\Gamma\left(\tfrac{d+3}{2}\right)\zeta(d+1)}\right]^{\tfrac{1}{d-1}}.
\eeq{a8} Similar relations show up in the brick wall model, propounded in~\cite{brickwall}, and subsequently used for $\text{dS}_3$
in~\cite{brickwall1}. Note that the cutoff $T_\text c$ is independent of the Hubble parameter $H$, as expected, but depends on the number
of DOFs in a way that is in complete accordance with the results of~\cite{Gia}. With some reasonable assumptions on $\mathfrak D$,
the right-hand side of Eq.~(\ref{a8}) is $\mathcal O(1)$. This sets $T_\text c\sim M_\text P$. Then, it is easy
to see that the thickness of the stretched horizon is $\mathcal{O}(l_{\text{P}})$.
With the identification~(\ref{a8}), one can now make explicit the area dependence of the entropy: \beq S=\frac{A}{4}\left(
\frac{d-1}{d}\right)(1-\epsilon)^{d/2}\,_2F_1\left(\tfrac{1}{2},\,1;\,\tfrac{d+2}{2};\,1-\epsilon\right),\eeq{entropyarea}
where the parameter $\epsilon$ depends on $A$ as follows \beq \epsilon=\frac{1}{4\pi^2}\left[\frac{8\,
\mathfrak{D}\,\Gamma\left(\tfrac{d+3}{2}\right)\zeta(d+1)}{A\,(d-1)\,\pi^{3/2}\,\Gamma\left(\tfrac{d}{2}\right)}
\right]^{\frac{2}{d-1}}.\eeq{epsarea} In the limit $\epsilon\rightarrow0$ or $A\rightarrow\infty$, thanks to Eq.~(\ref{a9}),
$S/A$ reaches the value $\tfrac{1}{4}$ for \emph{any} space dimensionality.

Similarly, the PI can define the total ``energy'' of the causal patch as follows. \beq E=\omega(d)\int_0^{r_\text c}
\frac{dr\,r^{d-1}}{\sqrt{1-H^2r^2}}\;\rho(r),\eeq{energy} where $\rho(r)=\mathfrak D\,a(d)\,[T(r)]^{d+1}$ is the local energy
density that a FIDO at a radial position $r$ reports to the PI. Again, using the integral representation~(\ref{hypergeo})
one finds
\beq E=\left(\frac{H}{2\pi d}\right)\mathfrak D\,\omega(d)\,a(d)\left(\frac{\sqrt{1-\epsilon}}{2\pi}\,\right)^d
\left(\frac{1}{\sqrt \epsilon}\right)^d.\eeq{energy1}
Expressing $\sqrt\epsilon$ in terms of the cutoff temperature $T_c$ through the relation~(\ref{epsdef}), one arrives at a rather
counter-intuitive conclusion: the ``energy'' scales like $(1/H)^{d-1}\sim A$. This is due to the extra factor of $H$ appearing in
Eq.~(\ref{energy1}). More explicitly,
\beq E=\left(\frac{A}{2\pi d}\right)\mathfrak D\,a(d)\left(1-\epsilon\right)^{d/2} T_{\text c}^{\,d}.\eeq{energy2}
Finally, using the expression~(\ref{a8}) for $T_{\text c}$, one finds after some simplifications that
the total ``energy'' is given by \beq E=\frac{A}{4}
\left(\frac{d-1}{d+1}\right)(1-\epsilon)^{d/2}\left[\frac{(d-1)(\sqrt{\pi}\,)^{d+3}}{4\,\mathfrak{D}\,\Gamma\left(\tfrac{d+3}
{2}\right)\zeta(d+1)}\right]^{\tfrac{1}{d-1}},\eeq{energyarea}
That the total ``energy''~(\ref{energyarea}) follows an area law just like the total entropy was also noticed in Ref.~\cite{Padi}
for a box of ideal gas kept near a de Sitter horizon. It is therefore natural to consider the quantity $E$ as an attribute of
the de Sitter horizon.

Our total ``energy'' $E$ is unique and well defined in the following sense. As soon as
the de Sitter entropy~(\ref{entropyarea}) is taken to be finite, we must forgo the symmetry of different causal
patches~\cite{dSsymmetry}. In the \emph{given} causal patch, all the FIDOs are on equal footing in that they all follow
time-like trajectories, are in causal contact with one another, and of course experience the same causal horizon. Any
quantity to be attributed to the entire causal patch or to the horizon itself must not depend on which FIDO, be her the
PI or not, is assigned the job of defining it. In other words, all the FIDOs must agree upon the value of any such quantity.
Now that any FIDO can learn about the results of local measurements performed by any other FIDO, they all will have identical
sets of data for the density distributions $\sigma(r)$ and $\rho(r)$, and therefore will agree upon their respective volume
integrals $S$ and $E$. This is not the case if in the definition~(\ref{energy}) one inserts a redshift factor (as was
suggested in Ref.~\cite{Padi}), which itself depends on the position of the FIDO assigned to define the quantity.

Eqs.~(\ref{entropyarea}) and~(\ref{energyarea}) can be viewed as relations among three extensive properties of the horizon,
namely $S$, $E$ and $A$. Dividing Eq.~(\ref{entropyarea}) by~(\ref{energyarea}), one can also write
\bea \frac{S}{E}=\left(\frac{d+1}{d}\right)\left[\frac{4\,\mathfrak{D}\,\Gamma\left(\tfrac{d+3}{2}\right)\zeta(d+1)}
{(d-1)(\sqrt{\pi}\,)^{d+3}}\right]^{\tfrac{1}{d-1}}\nonumber\\ \times\,{}_2F_1\left(\tfrac{1}{2},\,1;\,\tfrac{d+2}{2};\,
1-\epsilon\right).\eea{a10}
It is clear, in view of Eq.~(\ref{a9}), that in the limit $\epsilon\rightarrow0$, the ratio
$S/E$ is finite, although both $S$ and $E$ diverge. For $\epsilon\ll1$, Eq.~(\ref{a10}) can be approximated as
\beq S\approx\left(\frac{d+1}{d-1}\right)\left[\frac{4\,\mathfrak{D}\,\Gamma\left(\tfrac{d+3}{2}\right)\zeta(d+1)}{(d-1)
(\sqrt{\pi}\,)^{d+3}}\right]^{\tfrac{1}{d-1}}\left(\frac{E}{M_{\text P}}\right).\eeq{last} The virtue of Eq.~(\ref{last})
is that it allows one to compare the entropies of de Sitter spaces with different $d$, for a given $E$ in Planckian units.
It does not make sense to compare the horizon areas of two spaces with different dimensionalities, and indeed $A$ does not
appear in~(\ref{last}).

\section{Is 4D Spacetime Entropically Favored?}

Let us first note that the Hawking/Unruh radiation is (approximately) thermal. Then, in order for the density distributions
$\sigma(r)$ and $\rho(r)$ to be smooth, the constituent particles of the radiation cannot have mass $m\gtrsim H$.
This also means that the accessible number of DOFs does not depend on $r$. If the Hubble parameter $H$ is smaller
than the mass of the lightest massive DOF, only \emph{strictly} massless particles would contribute to the DOFs constituting the
radiation. Known difficulties with massless higher spins~\cite{HS} then make it natural to consider only particles of spin $s\leq2$.
If there is a massless spin-1 particle, no other particle can be charged under this,
because otherwise interactions would render the radiation non-thermal. While the Hawking/Unruh effect is very fundamental and takes
place in all dimensions, a massless chargeless fermion can exist only in some particular dimensions. This rules out spin-1/2 and spin-3/2
particles as non-generic. Scalars are also ruled out, since there is no symmetry to assure their masslessness. So we are
left only with spin-1 and spin-2 particles, whose masslessness can be guaranteed by gauge invariance. Now, there is one and just one
massless spin 2, namely the graviton~\cite{HS}. More than one vector particle is not a possibility, because they either
confine and cease to exist as long-range particles (when they are mutually charged), or there is no way to distinguish them as different
constituent particles of the radiation. At any rate, that photons and gravitons could be the sole constituents of the radiation
may not seem so surprising given that these are the natural DOFs to consider at low energy.

Thus one can assume, quite justifiably, that for a small cosmological constant the Hawking/Unruh radiation will be a gas of photons and
gravitons, whose interactions are negligible. In $(d+1)$ spacetime dimensions photon has $(d-1)$ DOFs, while graviton has $\tfrac{1}{2}(d+1)(d-2)$.
The total number of DOFs is therefore
\beq \mathfrak D~\equiv~\mathfrak D(d)=(d-1)+\tfrac{1}{2}(d+1)(d-2).\eeq{dof}

Let us consider universes that evolve from some initial state into a final state of a de Sitter space with a small cosmological constant,
like our Universe~\cite{dS}. The asymptotically de Sitter universes are assumed to have different dimensionalities, but the same
values of the fundamental constants, which all can be set to unity (we do not consider scenarios in which the Planck mass
$M_{\text P}$ may depend on spacetime dimensionality). Now for any given value of the characteristic quantity $E$, one can use
Eq.~(\ref{last}) to formally consider $S$ as a function of the number of space dimensions $d$. Is there any particular value of $d$ that
is favored entropically?



In the regime $2\leq d\leq10$, when $d$ is treated as a continuous variable, its function $S/E$ has an absolute
maximum at $d\approx2.97$. An upper bound on $d$ is essential since the function increases monotonically as
$S/E\sim\sqrt{d/2\pi e}$ for large $d$. Explicitly, the respective values for $d=2, 3, 4,\dots10$ are
approximately $1.096, 1.495, 1.458, 1.425, 1.409, 1.404, 1.407, 1.414$, and $1.425$. For a given value of the
``energy'' $E$, the entropy is therefore maximum for $d=3$. In other words, a universe whose final configuration
is a de Sitter phase with a small cosmological constant is entropically favored to have three spatial dimensions.



\section*{Acknowledgement}
We are grateful to G.~Gibbons, C.~Hull, A.~Iglesias, E.~Joung, M.~Kleban, D.~Klemm, L.~Lopez, M.~Porrati, A.~Sagnotti, L.~Sorbo
and M.~Taronna for useful discussions. AM and RR are thankful for the kind hospitality respectively of the HECAP section of ASICTP,
Trieste, where part of this work was done, and the Centre of Theoretical Physics at Tomsk State Pedagogical University during the
conference QFTG'14. At the time of this work, RR was supported in part by Scuola Normale Superiore, by INFN and by the ERC Advanced
Investigator Grant no.\,226455 ``Supersymmetry, Quantum Gravity and Gauge Fields" (SUPERFIELDS).
\end{multicols}

\end{document}